# Magnetic behavior of new compounds, $Gd_3RuSn_6$ and $Tb_3RuSn_6$

**Sanjay K Upadhyay, Kartik K Iyer and E.V. Sampathkumaran**
*Tata Institute of Fundamental Research, Homi Bhabha Road, Colaba, Mumbai 400005, India*

**Abstract**

We report temperature ($T$) dependence of dc magnetization, electrical resistivity ($\rho$) and heat-capacity of rare-earth (R) compounds, $Gd_3RuSn_6$ and $Tb_3RuSn_6$, which are found to crystallize in the $Yb_3CoSn_6$-type orthorhombic structure (space group: *Cmcm*). The results establish that there is an onset of antiferromagnetic ordering near ($T_N$=) 19 and 25 K respectively. In addition, we find that there is another magnetic transition for both the cases around 14 and 17 K respectively. In the case of the Gd compound, the spin-scattering contribution to $\rho$ is found to increase below about 75 K as the material is cooled towards $T_N$, thereby resulting in a minimum in the plot of $\rho(T)$ unexpected for Gd based systems. Isothermal magnetization at 1.8 K reveals an upward curvature around 50 kOe. Isothermal magnetoresistance plots also show interesting anomalies in the magnetically ordered state. There are sign reversals in the plot of isothermal entropy change versus $T$ in the magnetically ordered state, indicating subtle changes in the spin-orientations with $T$. The results reveal that these compounds exhibit interesting transport properties.

# I. INTRODUCTION

It is a well-known fact in the field of strongly correlated electron systems that 4f hybridization strength plays the most decisive role on the magnetic behavior of Ce containing materials. Ferromagnetic Kondo lattices of Ce are less abundant compared to antiferromagnetic ones due to the tendency of ferromagnetic interaction to destroy the Kondo effect [1]. Therefore, the discovery of ferromagnetic Kondo compounds has been of constant interest. Recently, Gribanova et al [2] reported the formation of a new Ce based compound, $Ce_3RuSn_6$, and its magnetic susceptibility ($\chi$), electrical resistivity ($\rho$) and heat-capacity ($C$) properties. This compound was established to form in $Yb_3CoSn_6$-type orthorhombic structure (space group: *Cmcm*). It was found that this Ce compound undergoes ferromagnetic ordering below ($T_C$= 3K) with a strong influence of the Kondo effect. In order to understand the role of 4f-hybridization on Ce magnetism, it is important to understand the magnetic behavior of other heavy rare-earth (R) members of this family (that is, those in which 4f orbital has been known to be localized) to explore whether there is any breakdown of de Gennes scaling. We were therefore motivated to attempt to synthesize Gd and Tb analogues in this ternary family and to explore its magnetic properties. This is the primary aim of this work.

Gribanova et al [Ref. 2] presented the crystallographic features for the Ce analogue at length. We reproduce here essential features. The crystal structure, when projected on yz-plane is made up of alternating infinite slabs of Ce and Sn of $AlB_2$-type, interconnected via tetragonal antiprisms with Ru inside these antiprisms. In the Ce-Sn slab, the Sn atoms are shifted from the centers of the corresponding trigonal prisms and thus these prisms are distorted. The atomic arrangements result in two sites for Ce and Ce-Ru distances are unusually large (~3.3 to ~3.6 Å). The coordination for Ce at *4c* site is 11 and the nearest coordination sphere $RuSn_{10}$ can be described as a distorted tetragonal prism capped with three more atoms. The coordination number for Ce at *8f* site is 12 and its polyhedran $Ru_2Sn_{10}$ can be described as pentagonal prisms with two additional atoms.

# II. EXPERIMENTAL DETAILS

We synthesized two batches of polycrystalline samples for R= Gd and Tb by arc melting together several times stoichiometric amounts of high purity (>99.9%) constituent elements in an atmosphere of argon. The molten ingots were cut for measurements. Powder x-ray diffraction (XRD) patterns (obtained with Panalytical XPert MPD, Cu $K_\alpha$) for both cases (see figure 1) resemble that of Ce analogue [2], establishing the absence of any other phase within the detection limit of XRD technique (<2%). The specimens were examined by scanning electron microscope (SEM) (Ultra Field Emission SEM of Zeiss). The SEM images showed two regions with slightly different darkness and the energy dispersive x-ray analysis revealed that this is due to slight differences in stoichiometry. This finding is in agreement with Ref. 2. One batch in each case was subjected to annealing at $700^0$ C for 10 days and we noted that the fraction of darker region was found to be diminished drastically. Thus annealing was found to result in more homogeneous material with a composition closer to that of starting composition. However, we did not find much difference in the magnetic features of these two batches. The data presented in this article is restricted to those obtained on annealed ingots. Temperature ($T$) and magnetic-field ($H$) dependent dc magnetization ($M$) studies were carried out on a specimen of mass about 10 mg with the help of a commercial SQUID magnetometer (Quantum Design, USA). Incidentally, we noted that the conclusions obtained on powder specimens glued with GE-varnish are essentially the same as that obtained on the broken pieces of ingots. Therefore, possible role of texture effects on the conclusions obtained from the magnetization studies on the broken pieces of ingot is ruled out.

Heat-capacity and electrical resistivity studies (also in the presence of externally applied magnetic field) in the range 1.8-300 K were carried with a commercial Physical Properties Measurements System (Quantum Design, USA). Typical mass of the specimens for heat-capacity measurements is 5 mg and the dimensions of the pieces for resistivity studies are of length 1.5mm and area of cross-section of 0.5mm×2mm. In general, most of the studies were performed for the zero-field-cooled condition (ZFC, from 300 K) of the specimen, unless stated explicitly.

### III. RESULTS AND DISCUSSIONS

#### A. Dc magnetization

Figure 2a shows dc $\chi$ of $Gd_3RuSn_6$ measured in a magnetic field of 5 kOe below 40 K. There is a gradual increase in $\chi$ with decreasing temperature down to about 20 K; there is a peak at ($T_N=$) 19 K indicating the onset of magnetic ordering (presumably antiferromagnetic). A careful look at the data measured in the presence of 100 Oe (Fig. 2b) reveals that there is in fact another peak at 14 K, which is smeared by the application of 5 kOe. Clearly, there are two magnetic transitions, one at ($T_{N1}=$) 19 K and the other at ($T_{N2}=$) 14 K. It is not clear whether Gd ions at both the sites order magnetically at different temperatures, in which case these two transitions arise from different sites of Gd ions; alternatively, there is a change in the magnetic structure with varying temperature. There is an upturn well below 10 K, and from this data alone it is difficult to claim whether this is intrinsic to this material or due to an impurity phase. In the curves obtained in 100 Oe, the bifurcation of the curves obtained for zero-field-cooled and field-cooled (FC) conditions is rather weak below $T_N$, and the FC curve keeps increasing with decreasing $T$ well below $T_N$ (instead of remaining flat). These features are not typical of spin-glass freezing. In the paramagnetic state, the plot of inverse $\chi$ is linear down to about 65 K (shown for $H = 5$ kOe in figure 3a) and the effective moment ($\mu_{eff}= 8\pm0.05$ $\mu_B$/Gd) obtained from the Curie-Weiss fit is consistent with trivalent Gd. The paramagnetic Curie temperature ($\theta_p$) is about -40 ($\pm2$) K and the sign suggests dominance of antiferromagnetic correlations. It should be noted that the value of $T_N$ is far less than that of $\theta_p$, thereby indicating the existence of some degree of magnetic frustration. The origin of this frustration presumably lies in crystallographic disorder, which is also inferred from the $\rho(T)$ data (see below). We would like to add that there is no magnetic moment on Ru ions, as inferred from the $\chi(T)$ behaviour of the Y analogue, which is found to be diamagnetic with a temperature independent $\chi$ of about -0.00017 emu/mol (equivalent to $0.002 \times 10^{-6}$ m$^3$/mol).

Corresponding dc $\chi$ data are presented in figures 2c and 2d and inverse $\chi$ behavior is also shown in figure 3b for the Tb compound. It is clear that there is a shoulder and peak in the dc $\chi$ data in figures 2c and 2d near 25 and 17 K, which can be labelled as $T_{N1}$ and $T_{N2}$ respectively. There is also an upturn at a lower temperature (below about 8 K) as in the case of the Gd compound. The inverse $\chi$ plot (figure 3b) is linear over a wide $T$-range above $T_{N1}$ and the value of $\mu_{eff}$ obtained from Curie-Weiss fitting turns out to be 10 ($\pm0.05$) $\mu_B$/Tb, in good agreement with the theoretical value for trivalent Tb ion; the value of $\theta_p$ is about -26 ($\pm1$) K. It is interesting that, unlike in Gd case, the magnitude of $\theta_p$ is the same as that $T_{N1}$, as though magnetic frustration effect is absent in this case unlike in Gd. By the arguments advanced for the Gd analogue above, the absence of magnetic frustration implies relatively less crystallographic disorder. The relative values of residual resistivity ratios (presented below in this article) are consistent with this line of argument. The shapes of ZFC-FC curves obtained in 100 Oe look similar with insignificant bifurcation below $T_{N1}$, ruling out spin-glass freezing.

Figures 4(a-c) and 4(d-f) show $M(H)$ plots up to 70 kOe at some selected temperature (1.8, 10 and 16 K for Gd case and 1.8, 5 and 22 K for Tb case) below $T_N$. There is no evidence for a tendency towards saturation even at higher fields (with measurements extended up to 140 kOe) for both the compounds. This finding is consistent with antiferromagnetism in the two *T*-ranges $T_{N1}>T>T_{N2}$ and $T<T_{N2}$. The plots are looking similar in the magnetically ordered states, suggesting that antiferromagnetism prevails following magnetic ordering of R ions at all sites. A careful look at the curves, say at 1.8 K, reveals that there is an upward curvature around 30-60 kOe, indicating field-induced effects in the spin-reorientation. The curves are almost non-hysteretic.

### B. Heat capacity

In order to get further support for the magnetic ordering behavior, we show the heat-capacity behavior in figures 5a and 5d. The heat capacity curves are also presented in figures 5b and 5e in the form of *C/T vs. T*. It is clear that, in zero field, the heat-capacity exhibits features at the same temperatures as in χ(*T*), thereby confirming the existence of the two magnetic transitions. In the case of Tb compound, the anomaly at $T_{N1}$ is very weak. In the presence of external magnetic fields, the peaks shift towards lower temperatures, as shown in the insets of figures 5a and 5d; this is a strong support to antiferromagnetic nature of both the magnetic transitions.

We have also measured heat-capacity of the Y analogue to derive the magnetic contribution, $C_m$ to *C* and the curve obtained after subtracting the lattice part using the method suggested by Bouvier et al [3]. The curves related to lattice part are also shown in figures 5a and 5d. It is worth noting that the lattice part tends to merge with the experimental data well beyond $T_{N1}$ in the case of Gd system, indicating good estimation of lattice contribution. However, in the case of Tb compound, deviation between the experimental data and the (derived) lattice part persists well beyond $T_{N1}$; this is attributed to possible existence of a broad Schottky anomaly above $T_{N1}$, which should be absent for the Gd case (for which orbital angular momentum is zero). The magnetic entropy, $S_m$, derived from $C_m$ are shown in the insets of figure 5b and 5e. In the case of Gd compound, the value of $S_m$ at $T_{N1}$ is about 16 J (Gd mol)$^{-1}$ K$^{-1}$, which is marginally lower than the theoretically expected value of 17.3 J mol$^{-1}$ K$^{-1}$. This lower value presumably arises from the persistence of short-range magnetic correlations well above $T_N$. For the Tb compound (see figure 5e), the value of $S_m$ at $T_{N1}$ is about 12 J (Tb mol)$^{-1}$ K$^{-1}$, which is less as compared to the value for the fully degenerate trivalent Tb ion (21.3 J mol$^{-1}$ K$^{-1}$); since crystal-field effects can also dominate in the Tb case, it is not possible for us at present to delineate the effects due to possible short-range magnetic order.

### C. Magnetocaloric behavior

In order to infer about magnetocaloric behaviour, we have derived isothermal entropy change [*ΔS = S(H)-S(0)*] on the basis of *C(T)* data obtained in the presence of external fields. For the Gd case, the values are found to be negligible for a change of *H* to 10 kOe from zero, but gets significant (though still small) for higher values of *H*, as shown in figure 5c. The sign of –*ΔS* is positive as $T_{N1}$ is approached by decreasing the temperature resulting in a peak at about 20 K for the final fields, *H*= 50, 100 and 140 kOe; this positive peak [4] suggests magnetic-field-induced spin-reorientation resulting in a ferromagnetic component. Such a weak peak is found to be present even for *H*= 30 kOe, though the magnitude at the peak is very small. The values of –*ΔS* tend to become negative below $T_N$, in agreement with the fact that the low-field state is antiferromagnetic. Judged by the observation of another positive peak in -*ΔS* around 5 K, the presence of a ferromagnetic component is inferred at such low temperatures; it is therefore possible that the low-temperature tail in χ may also be intrinsic due to subtle spin-reorientation effects. The features in *ΔS* for the Tb case (see figure 5f) are in some respects similar to that of Gd case, e.g., in sign-crossover. A (broad) positive peak of -*ΔS* can be seen at $T_{N1}$ for *H*>10 kOe and at $T_{N2}$ for a given $H$, sign of $-\varDelta S$ becomes negative. Beyond, the spin-reorientation transition field, say, for $H=100$ Oe, positive $-\varDelta S$ can be seen.

### D. Electrical resistivity and magnetoresistance

The temperature dependencies of electrical resistivity in the absence as well as in the presence of magnetic field are shown in the figures 6 and 7. Insets show the data obtained in the absence of magnetic field in an expanded form below 40 K to highlight the features due to magnetic ordering. The values of ρ need to be viewed with some degree of caution due to the difficulties arising out of brittleness of the specimen, as noted in Ref. 2.

Let us first discuss the results for Gd case. It is to be noted that the residual resistivity ratio, $\rho(300\ K)/\rho(1.8K)$ is close to 1.009. This small ratio signals significant crystallographic disorder. However, some qualitative inferences can be made from this data: (i) In the zero-field curve, following a positive coefficient of ρ down to about 75 K, there is an upturn with decreasing temperature resulting in a minimum (not expected for Gd systems), mimicking the behaviour in the Ce analogue [2]. This minimum can not arise from antiferromagnetic energy-gap formation, as antiferromagnetism sets in at a much lower temperature. Since the Kondo effect is not possible in such a strictly localized 4f system, we believe that this upturn is due to increased spin-scattering as a consequence of crystallographic disorder, as a precursor to long-range magnetic order. Such an interesting finding has been occasionally reported for some Gd systems in the past as well by us [5]. As proposed by us in Ref. 5, there is a possibility that conduction electrons polarized by s-f exchange interaction gets trapped as the temperature is lowered towards $T_{N1}$; just as crystallographic disorder causes electron-localization, short-range magnetic order could facilitate the proposed electron localization process. (ii) The drops due to spin-disorder contribution are not visible in the raw data neither at $T_{N1}$ nor at $T_{N2}$, but a broad peak (possibly as a result of superimposition of these two drops) appears around 17 K, followed by a drop at lower temperatures. (iii) There is an additional drop at about 6 K with decreasing $T$, which could be attributable to the existence of subtle spin-reorientation, mentioned earlier.

It is obvious from the figure 6a that the $\rho(T)$ minimum gets gradually suppressed with increasing field, resulting in its absence for $H=100$ kOe. The magnetoresistance [MR= $\rho(T)-\rho(0)]/\rho(0)$] is negative in the paramagnetic state with its magnitude increasing with decreasing temperature below ~100 K for a given field; the magnitude of isothermal MR also increases with increasing $H$ above $T_N$, as inferred from the figure 6a. The peak temperature due to the onset of magnetic order in the plot of $\rho(T)$ undergoes a gradual suppression with increasing $H$, for instance to ~14K in 100 kOe, consistent with the findings from $H$-dependence of $C$. We have collected isothermal MR data at selected temperatures in the magnetically ordered state, say at 1.8, 10 and 16 K (see figure 6 b-d). The magnitudes of MR in general are not very large. However, some qualitative changes with increasing temperature can be seen. Arrows and numericals are placed on the curves as a guide to follow the curves while changing the magnetic field. Looking at the 1.8 K data (figure 6b), it is apparent that there is a sharp change for small applications of magnetic field. However such a feature can not be seen in the $M(H)$ curve (see figure 4a); in fact, we obtained the $\chi(T)$ curve in a low field of 30 Oe and it looks the same as that obtained in 100 Oe. We therefore believe that the initial sharp change in MR in the virgin curve for this temperature may arise from dominating scattering effect at the grain boundaries which are influenced by low fields. The sign of MR is positive, consistent with dominant antiferromagnetism and its magnitude increases with increasing $H$, exhibiting a peak around 30 kOe. The sign of MR gets reversed near 50 kOe, consistent with the conclusion from isothermal M curves that there is a magnetic-field induced changes in the magnetic structure. On reversing the field towards zero and on further cycling in magnetic field, an envelope curve is obtained and it is found that the virgin curve lies outside this envelope curve. This implies that the magnetic-field induced change is first-order in nature, however broadened by disorder [6]. At 10 and 16 K (figures 6c and 6d respectively), the positive peak in MR is weakened; the above-mentioned virgin-curve behaviour is absent and reversible MR curves are obtained.

Now, turning to the Tb case, negative temperature coefficient of resistivity interestingly is not observed above $T_{N1}$ (see figure 7a), unlike the situation encountered for the Gd case. Residual resistivity ratio is much larger (about 2.5) compared to the Gd case, as though crystallographic disorder is comparatively less in this case. There is a drop in ρ at the onset of magnetic ordering; the temperature derivative of ρ (not shown in the form of figure) reveals a change of slope around $T_{N2}$. The sign of MR is negative and the magnitude is small in the paramagnetic state (e.g., about 1% even for $H$= 70 kOe at 40 K, see figure 7d) in the paramagnetic state. However, well below $T_N$, the sign of MR is positive and the magnitude gradually gets larger, as brought out in the figure 7b for $T$= 1.8 K, without any sign reversal (observed for the Gd case). Though the absence of sign reversal even at high fields following spin alignment is not clear to us at present, positive magnetoresistance even in ferromagnetic systems has been reported occasionally in the literature [7]. Figure 7b also reveals that the virgin curve lies outside the envelope curve, as though there is a broadened first order transition with the application of $H$ in this compound as well. However, this peculiar behaviour is not seen in the $M(H)$ curve (Fig. 7d). The reason for this discrepancy is not clear at this moment. Between $T_{N1}$ and $T_{N2}$, say, at 22 K (see figure 7c), there is a (positive) peak near 40 kOe, followed by sign reversal at higher fields. This competition between positive and negative contributions indicates possible changes in Fermi surface contributions following spin-reorientation induced by external magnetic field.

### IV. SUMMARY

We have reported the formation of new rare-earth based compounds, $R_3RuSn_6$ (R= Gd and Tb) and studied their magnetic properties by magnetic susceptibility, electrical resistivity, magnetoresistance and heat capacity measurements. We find that these compounds order antiferromagnetically in the absence of magnetic field. There are at least two magnetic transitions in both these compounds. The long range magnetic ordering sets in at a higher temperature for the Tb compound (near 25 K), compared to the Gd case (near 19 K); though this is opposite to what one would expect on the basis of de Gennes scaling, it is not an uncommon finding [8] and is attributed to the role played by anisotropic 4f orbitals following crystal-field splitting for the Tb case. For instance, the readers may see Ref. 9 for a similar enhancement of magnetic ordering temperature for the Tb compound in the series, RPdSn, and for quantitative explanation taking into account crystalline electric-field term to the exchange Hamiltonian, showing how the sign of the simple $B^0_2$ term results in the maximum ordering temperature for the Tb compound in the RPdSn series.

According to de Gennes scaling, the magnetic transition for the corresponding Ce compound should be below 0.2 K. Instead, the Ce compound, $Ce_3RuSn_6$, has been reported to order at a rather higher temperature, that is, below 3 K. Clearly, there is a breakdown of de Gennes scaling for this Ce compound. Therefore, following previously established facts for Ce compounds in the literature, the enhanced magnetic ordering for $Ce_3RuSn_6$ arises from stronger 4f hybridization. There is an evidence for a broad magnetic-field-induced spin alignment below $T_N$ from isothermal $M(H)$ and MR data.

Finally, the Gd compound belongs to a small family of Gd compounds, in which there is a resistivity minimum around 75 K (in the paramagnetic state). This finding is noteworthy considering that the electrical resistivity of Tb analogue does not exhibit such a feature. Different degree of crystallographic disorder between these compounds could be responsible for such qualitative differences in their transport properties.

**REFERENCES**

1. See, for instance, A. Bhattacharyya, C. Ritter, D. T. Adroja, F. C. Coomer, and A. M. Strydom, Phys. Rev. B **52** (2016) 014418.
2. V. Gribanova, D. Gnida, E. V. Murashova, A. V. Gribanov, and D. Kaczorowski, J. Alloys and comp. **671** (2016) 114.
3. M. Bouvier, P. Lethuillier, and D. Schmitt, Phys. Rev. B **43** (1991)13137.
4. V. K. Pecharsky and K. A. Gschneidner Jr., J. Magn. Magn. Mater. **200** (1999) 44 and K. A. Gschneidner Jr., V. K. Pecharsky and A. O. Tsokol, Rep. Prog. Phys. **68** (2005) 1479.
5. R. Mallik, E. V. Sampathkumaran, M. Strecker and G. Wortmann. Europhys. Lett. **41**, (1998) 315 and E. V. Sampathkumaran, K. Sengupta, S. Rayaprol, and K. K. Iyer, Phys. Rev. Lett **91** (2003) 036603; R. Mallik and E.V. Sampathkumaran, Phys. Rev. B **58** (1998) 9178.
6. S. B. Roy and P. Chaddah, Phase Transitions **77** (2004) 767; P. Chaddah and S. B. Roy, Phys. Rev. B **60** (1999) 11926.
7. R. Mallik, E. V. Sampathkumaran and P. L. Paulose, Appl. Phys. Lett. **71** (1997) 2385.
8. See, for instance, K.A. Gschneidner, Handbook on the physics and chemistry of rare-earths, Elsevier, **2** (1979) and references therein.
9. D.T. Adroja and S.K. Malik, Phys. Rev. B **45,** 779 (1992).

**Figure 1:** Powder x-ray diffraction patterns (Cu Kα) of polycrystalline **(a)** $Gd_3RuSn_6$ and **(b)** $Tb_3RuSn_6$.

**Figure 2:** Dc magnetic susceptibility as a function of temperature below 40 K for $Gd_3RuSn_6$ measured in a magnetic field of **(a)** 5 kOe and **(b)** 100 Oe. Corresponding curves for $Tb_3RuSn_6$ are plotted in **(c)** and **(d)** respectively.

**Figure 3:** Inverse magnetic susceptibility as a function of temperature (1.8 – 300 K) for **(a)** $Gd_3RuSn_6$ and **(b)** $Tb_3RuSn_6$.

**Figure 4**: Isothermal magnetization at selected temperatures for **(a-c)** $Gd_3RuSn_6$ (1.8, 10 and 16 K) and **(d-f)** $Tb_3RuSn_6$ (1.8, 5 and 22 K). A weak irreversibility of *M(H)* can be seen for the Tb case below $T_N$ and, for 1.8 K, the virgin curve lies inside the loop.

**Figure 5:** **(a)** Heat-capacity and **(b)** heat-capacity divided by temperature for $Gd_3RuSn_6$. Corresponding curves for Tb analogue are plotted in figures **(d)** and **(e)** respectively. The heat-capacity data for the lattice part derived as mentioned in the text are also shown in (a) and (d). Insets in (a) and (d) show the heat-capacity data in the presence of external fields (*H*= 0, 5, 100 and 140 kOe for Gd case and 0, 30, 50, and 100 Oe for Tb case) in the vicinity of magnetic transitions. The magnetic entropy is plotted in the insets of figures (b) and (e). Isothermal entropy change for a variation of the magnetic field from zero is plotted in **(c)** and **(f)** for Gd and Tb compounds.

**Figure 6:** Electrical resistivity **(a)** as a function of temperature in zero field as well as in the presence of external fields for $Gd_3RuSn_6$. The data below 40 K is shown in an expanded form in the inset. Isothermal magnetoresistance behavior at 1.8, 10 and 16 K are shown in **(b)**, **(c)** and **(d)** respectively. In (b), arrows and numericals are placed on the curves as a guide to the eyes.

**Figure 7:** Electrical resistivity **(a)** as a function of temperature in zero field as well as in the presence of external fields (30 and 50 kOe) for $Tb_3RuSn_6$. The zero-field data below 40 K is shown in an expanded form in the inset. Isothermal magnetoresistance behavior at 1.8, 22 and 40 K are shown in **(b), (c)** and **(d)** respectively.

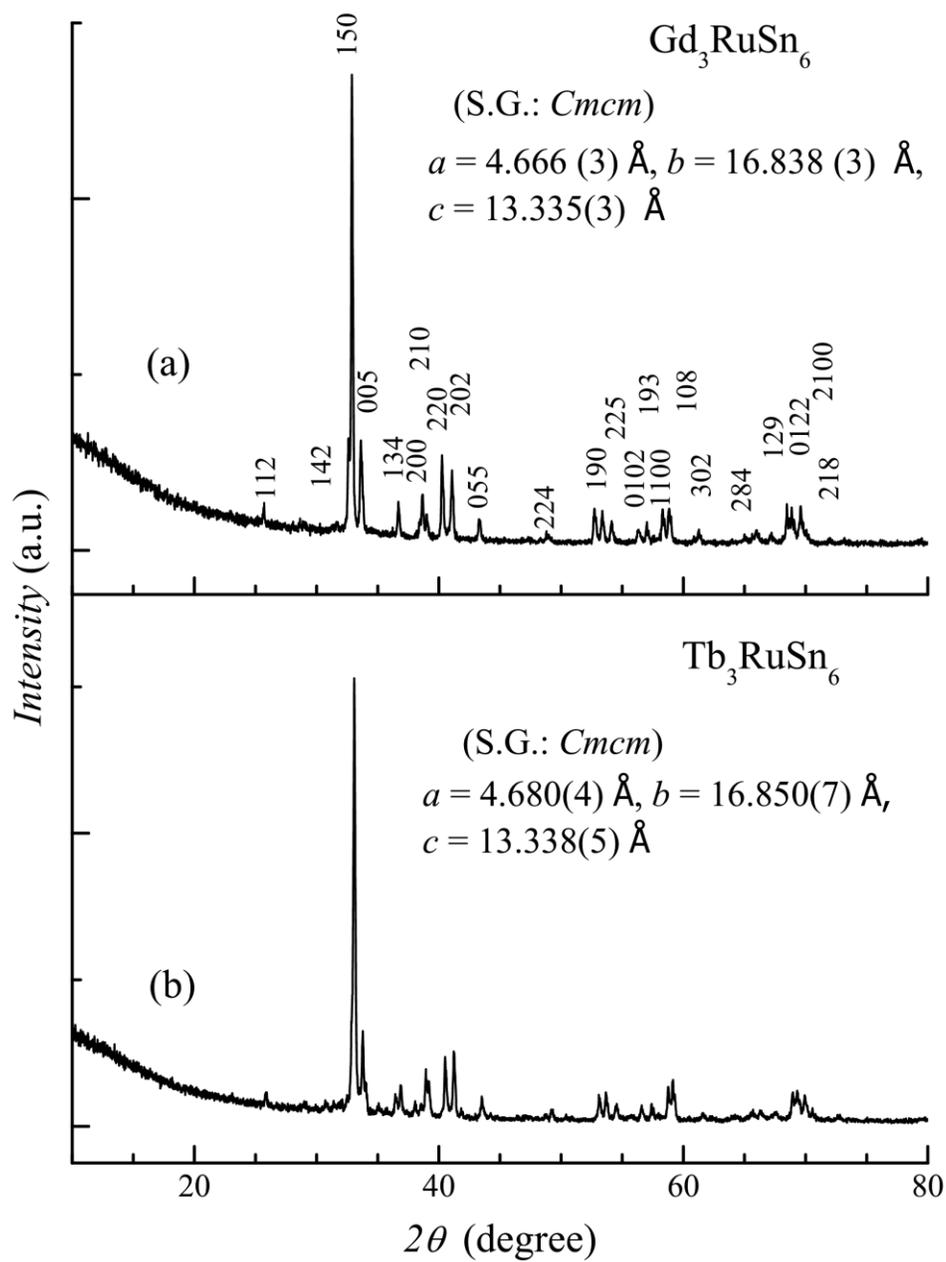

Figure.1

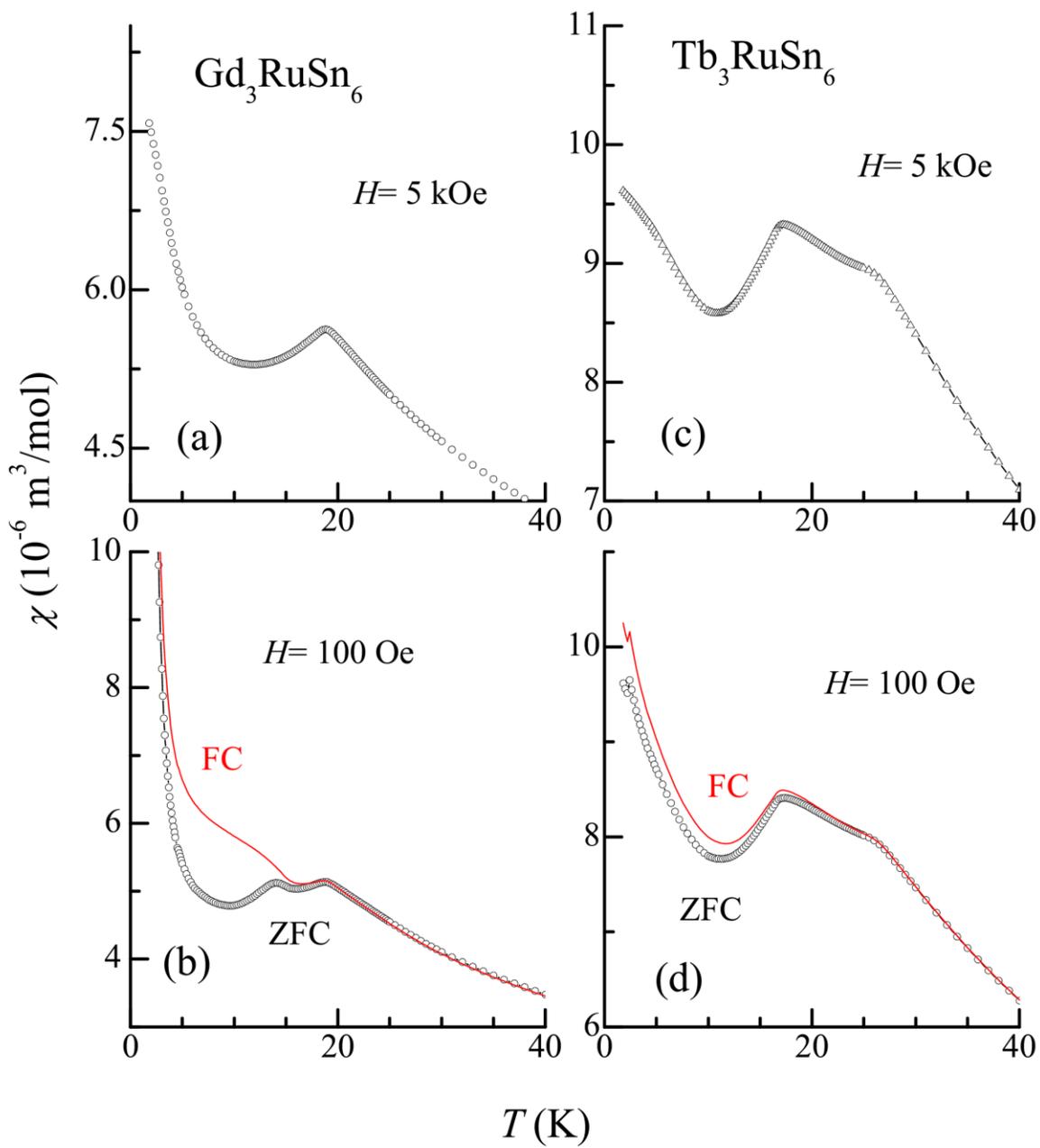

Figure.2

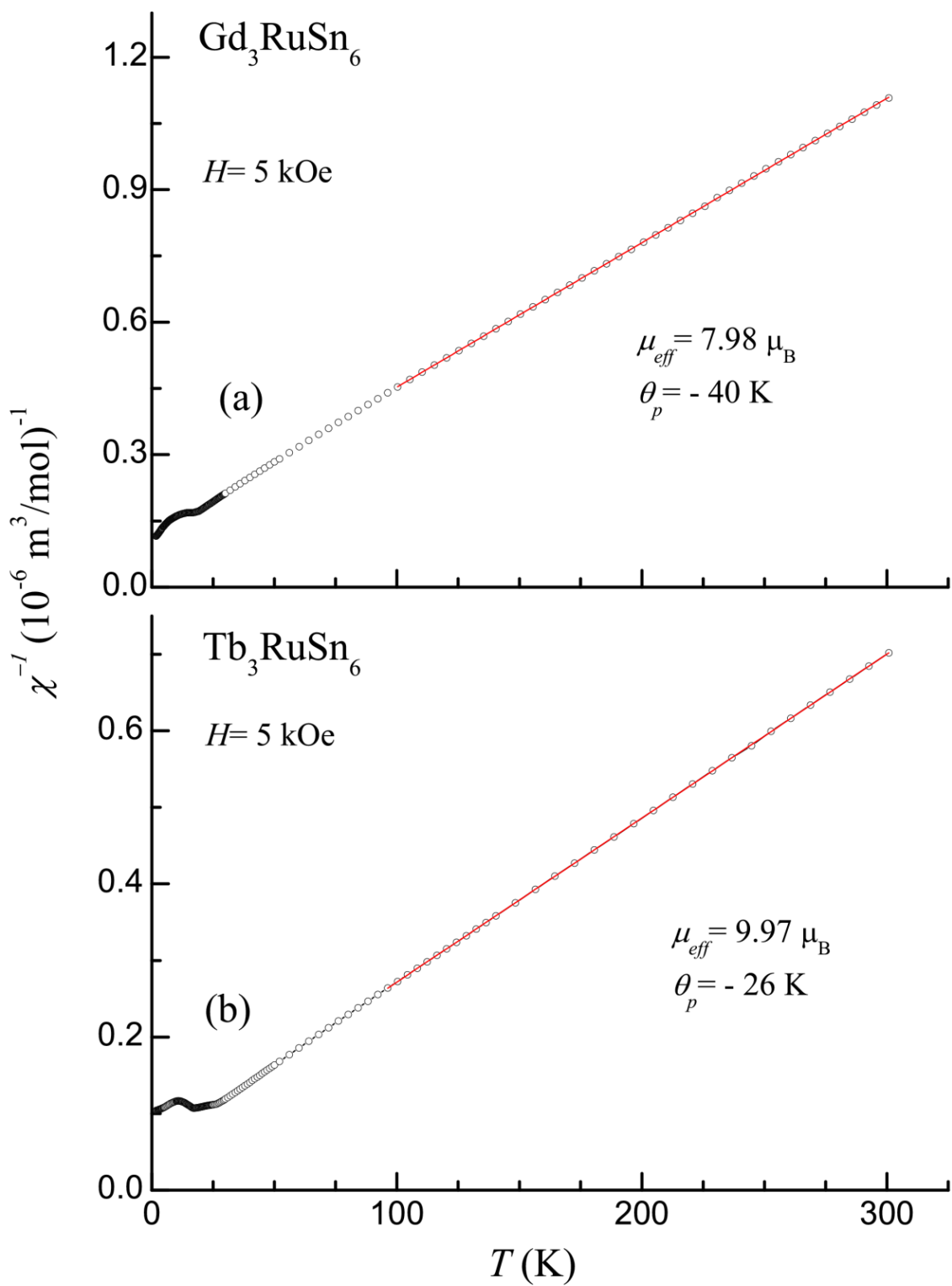

Figure.3

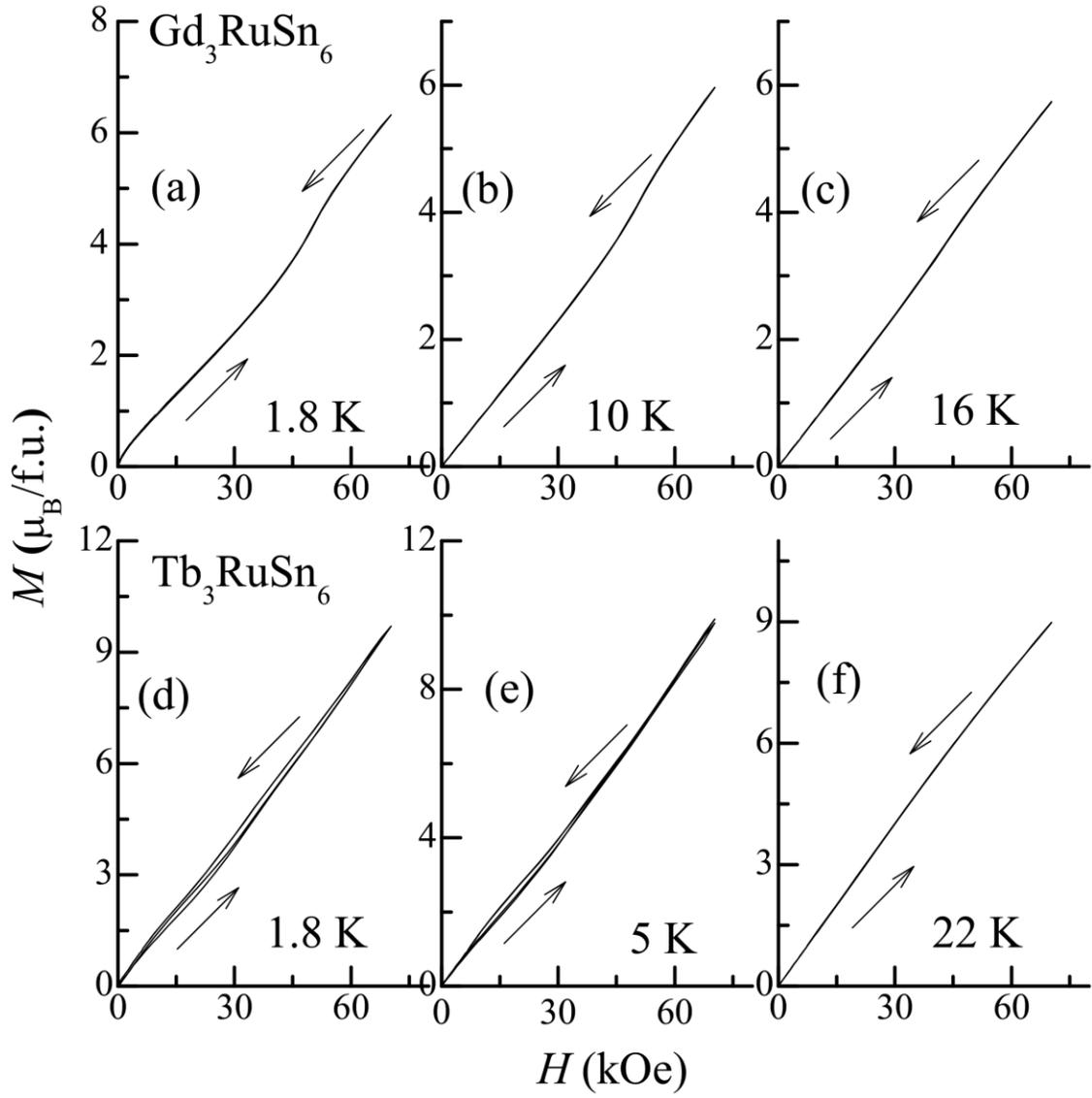

Figure.4

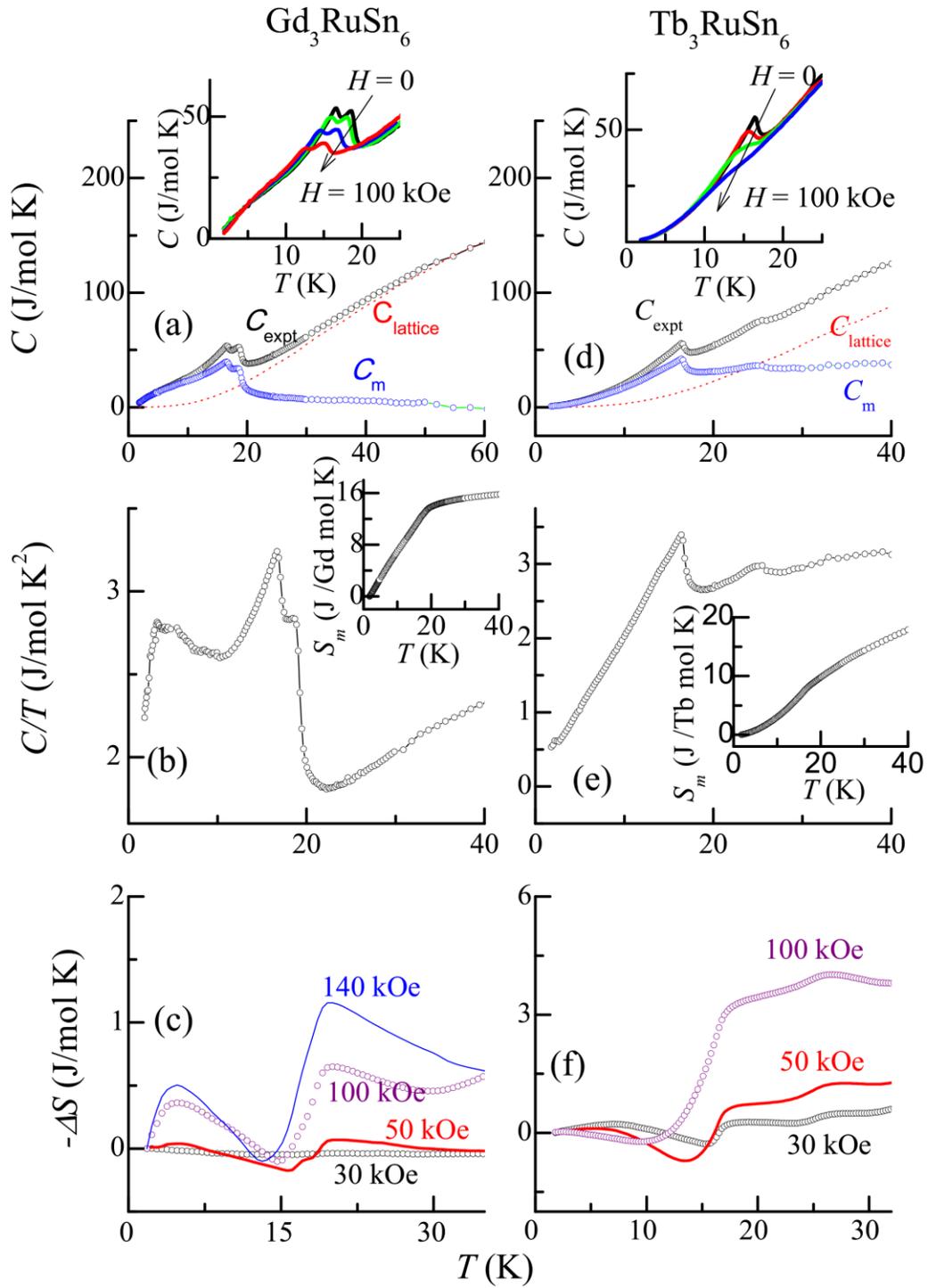

Figure.5

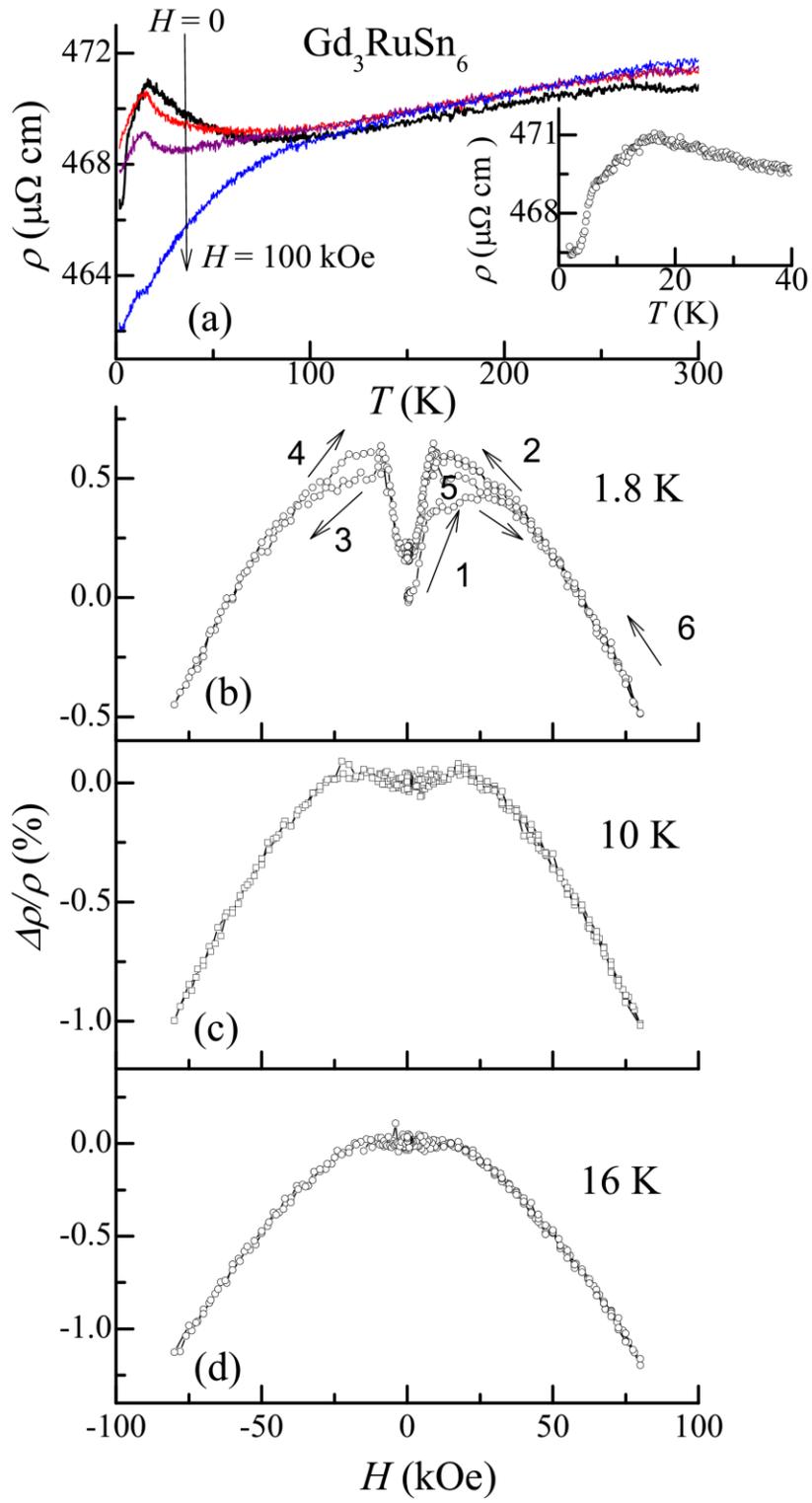

Figure.6

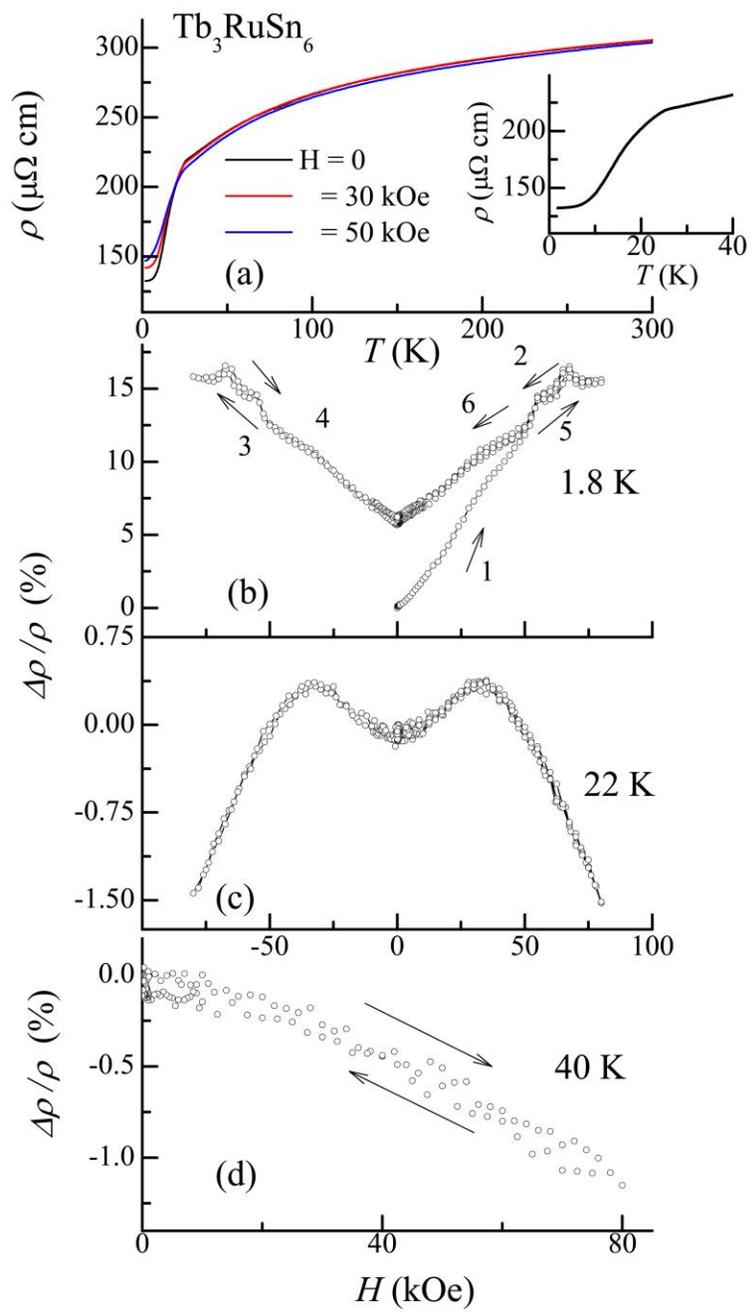

Figure.7